\documentclass[usegraphicx,usedcolumn,usenatbib]{mn2e}
\usepackage{graphicx}
\usepackage{txfonts}
\usepackage{subfigure}
\usepackage{rotating}
\usepackage{accents}

\newcommand{\apj}{ApJ}

\newcommand{\apjs}{ApJS}
\newcommand{\aj}{AJ}
\newcommand{\mnras}{MNRAS}

\newcommand{\aap}{A\&A}

\newcommand{\Msun}{M$_{\odot}$}
\newcommand*{\dt}[1]{ %
 \accentset{\mbox{\large\bfseries .}}{#1}}

\bibpunct[,]{(}{)}{;}{a}{,}{,}

\begin{document}

\title[Composition of a young star cluster ejecta]
{Modelling the composition of a young star cluster ejecta}

\author[Moll\'{a} \& Terlevich]
{Mercedes ~Moll\'{a} $^{1}$\thanks{E-mail:mercedes.molla@ciemat.es}
 and Roberto ~Terlevich $^{2,3}$ \thanks{Visiting Professor UAM, Madrid}\\ 
$^{1}$ Departamento de Investigaci\'{o}n B\'{a}sica, CIEMAT,
Avda. Complutense 40. E-28040 Madrid. (Spain)\\
$^{2}$ INAOE, Luis Enrique Erro 1, Tonanzintla, Puebla 72840, (Mexico)\\
$^{3}$Institute of Astronomy, University of Cambridge, Madingley Road, 
Cambridge, CB3 0HA, (UK)}

\date{Accepted Received ; in original form }

\pagerange{\pageref{firstpage}--\pageref{lastpage}} \pubyear{2011}

\maketitle \label{firstpage}

\begin{abstract}

We have computed with a fine time grid the evolution of the elemental
abundances of He, C, N and O ejected by young ($t < 20$\,Myr) and
massive (M $=10^{6}$\,\Msun) coeval stellar cluster with a Salpeter
initial mass function (IMF) over a wide range of initial abundances.
Our computations incorporate the mass loss from massive stars (M~$\rm
\ge 30\,M_{\odot}$) during their wind phase including the Wolf-Rayet
phase and the ejecta from the core collapse supernovae. We find that
during the Wolf-Rayet phase ($t < 5$\,Myr) the cluster ejecta
composition suddenly becomes vastly over-abundant in N for all initial
abundances and in He, C, and O for initial abundances higher than
1/5th Solar. The C and O abundance in the cluster ejecta can reach
over 50 times the solar value with important consequences for the
chemical and hydrodynamical evolution of the surrounding ISM.

\end{abstract}


\section{Introduction}

Massive young stellar clusters are ideal laboratories for research
into the evolution of massive stars and their interaction with their
surrounding interstellar medium (ISM). These luminous and rapidly
evolving massive stars supply most of the young cluster UV radiation that
creates the encompassing H{\sc ii} region and a large amount of mass
and mechanical energy in the forms of supernova ejecta and stellar
winds particularly during the WR phase. These massive starforming
regions can eject during their first 10 Myr of evolution about 20\% of
their initial mass \citep{stb99} (hereinafter STB99) mostly in the
form of newly synthesized C, N and O that by mass represent most of
the heavy elements.

Massive young stellar clusters are ubiquitous particularly among late
type galaxies. Their stellar wind phase can result in a supergalactic
wind affecting the nearby intergalactic medium (IGM).  It is open to
question how the ejected freshly synthesized heavy elements cools and
mix with the ISM, and how long this process governed by the cooling
time scale, may last. The time scale for cooling is strongly dependent
on the gas cooling rate that in turn is dependent on the gas chemical
composition and density
\citep[e.g.][]{ten96,kob97,kob98,vzee06,arls10}.  Thus evaluating the
composition of the cluster ejecta and its time evolution is a
necessary prior step for estimates of the evolution of the cooling
function and the computation of cooling and feedback time scales.
Although work like that of \citet{sil01} underlined the large
influence that the enrichment of the stellar ejecta can have on the
radiative cooling of starburst superbubbles, many researchers are still
using \citet{ray76} radiative cooling coefficient calculation for
Solar abundances when modelling the interaction between the stellar
ejecta and their surrounding medium.

In this paper we present a set of models designed to calculate in
detail the first 20\,Myr of the evolution and chemical composition of
a star cluster ejecta on very short time scales, i.e. much shorter
than the H{\sc ii} region lifetime and in particular to resolve the WR
wind phase

Galactic chemical evolution models traditionally assume that the
elements ejected by the stellar cluster are incorporated to the ISM
when the corresponding stars die, i.e. at a time equal to their
lifetimes \citep{pcb98,dray03,dray03b}. The shortest time step is usually
defined by the mean-lifetime of the most massive star.  Since this is
typically around 100--120\,\Msun, chemical evolution calculations
begin normally at around 3-5\,Myr with comparable time-steps. This way
important phases of the wind evolution, occurring before 3\,Myr or
short lived like the WR phase are lost or diluted.

To compute the composition of the ejecta, most chemical evolution
models use the total yields of elements due to supernova explosions, such as
those given by \citet[][WW95]{woo95} or other more recent works
\citep{ume02,rauscher02,lim03,chi04,fro06,heger10} to calculate the
total change of the elemental abundances. Other computations include
the elements ejected during the wind phase of massive stars
\citep{mae92,mey02,hir05,hir07,kob06}, only the models of
\citet[][hereinafter PCB98]{pcb98} include the evolution of both
phases, i.e.  the yields of core collapse supernova explosions from
\cite{woo95} and the stellar wind yields produced during the evolution
of each star. However, PCB98 computations, as most chemical evolution
models, were performed with time steps much longer than the lifetime
of an H{\sc ii} region therefore missing short lived stages like the
WR phase.  Moreover, the evolution of a supernova progenitor that
loses part or most of its mass is not the same as a normal main
sequence massive star. Since mass and structure are substantially
different in he time of the supernova explosion, the associated yields
will also differ, such as it is explained in \citet[][ hereinafter
WLW93 and WLW95, respectively]{wlw93,wlw95}. To take this into account,
PCB98 linked the final stage of the star after lost mass with the
supernova yields though the CO core mass. However, they use the WW95
yields instead WLW93/WLW95 yields.

The organization of this paper is as follows: In Section 2 we use O
and WR winds computations to estimate the evolution of the ejecta of a
young massive stellar cluster.  In section 3 we compute the
contribution of the explosive nucleosynthesis. Section 4 gives the
complete evolution of elemental abundances within the cluster and
discusses the impact of each contribution phase over the final
ejecta. Our conclusions are given in Section 5.

\section{Stellar wind composition}

\begin{figure*}
\subfigure{\includegraphics[angle=0,width=\textwidth]{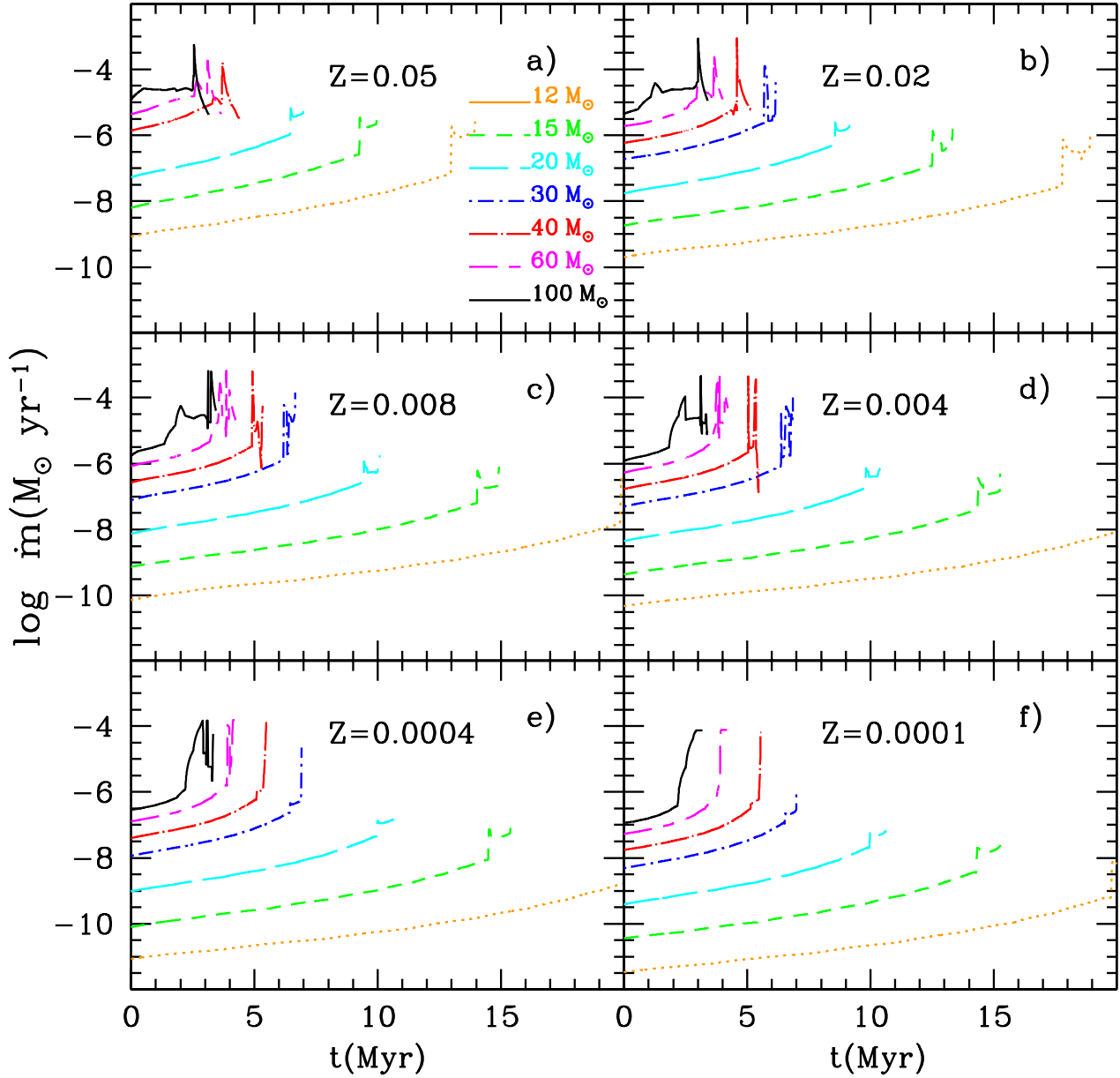}}
\caption{Evolution of the mass loss rate for the stars given by the
Padova group for different metallicities Z as labelled. Color and type
of lines mean a different value of stellar mass as labelled.}
\label{mdot_Z}
\end{figure*}

To compute the evolution of the rate of ejection of elements due to
stellar winds of a Single Stellar Population (SSP) of a given initial
metallicity and initial mass function (IMF), we have used the tables
resulting from the isochrones calculation by
\citet{bre93,fag94a,fag94b}.  These authors give the amount of mass
lost by massive stars during their evolution for 7 initial stellar
masses: 12, 15, 20, 30, 40, 60 and 100\,M$_{\odot}$. For each initial
mass and time step, the tables provide the present mass (in solar mass
units), the rate of mass loss in M$_{\odot}$ yr$^{-1}$ (logarithmic
scale) and the abundances of the stellar surface for H, $^{4}$He,
$^{12}$C, $^{14}$N, and $^{16}$O. These tables are provided for 6
initial metallicities: Z=0,0001, 0.0004, 0.004, 0.008, 0.02 and 0.05.

\begin{table}
\caption{Characteristics of the stellar models}
\begin{tabular}{lrrrr}
\hline
Z & m$_{*}$ & m$_{end}$ & t$_{end}$ &X$_{f,H}$\\
  & (M$_{\odot}$) & (M$_{\odot}$) & (Myr) &  \\   
\hline
 0.0001 &      12 &    11.99  &   21.17  &  0.748 \\        
 0.0001 &      15 &    14.98  &   15.28  &  0.731 \\        
 0.0001 &      20 &    19.93  &   10.65  &  0.701 \\         
 0.0001 &      30 &    29.69  &    7.00  &  0.770 \\         
 0.0001 &      40 &    38.93  &    5.52  &  0.770 \\         
 0.0001 &      60 &    37.31  &    4.19  &  0.681 \\        
 0.0001 &     100 &    67.39  &    3.18  &  0.649 \\         
 0.0004 &      12 &    11.97  &   21.39  &  0.741 \\         
 0.0004 &      15 &    14.94  &   15.39  &  0.727 \\        
 0.0004 &      20 &    19.83  &   10.64  &  0.699 \\         
 0.0004 &      30 &    29.35  &    6.91  &  0.673 \\          
 0.0004 &      40 &    35.66  &    5.46  &  0.693 \\          
 0.0004 &      60 &    31.89  &    4.20  &  0.486 \\         
 0.0004 &     100 &    55.45  &    3.32  &  0.142 \\          
  0.004 &      12 &    11.84  &   21.47  &  0.716 \\          
  0.004 &      15 &    14.76  &   15.26  &  0.705 \\          
  0.004 &      20 &    19.30  &   10.48  &  0.686 \\         
  0.004 &      30 &    19.05  &    6.86  &  0.544 \\          
  0.004 &      40 &    16.86  &    5.44  &  0.241 \\          
  0.004 &      60 &    17.55  &    4.20  &  0.000 \\         
  0.004 &     100 &    46.93  &    3.38  &  0.080 \\           
  0.008 &      12 &    11.77  &   21.07  &  0.708 \\         
  0.008 &      15 &    14.62  &   14.92  &  0.693 \\          
  0.008 &      20 &    19.03  &   10.10  &  0.671 \\         
  0.008 &      30 &    13.05  &    6.66  &  0.516 \\          
  0.008 &      40 &    16.44  &    5.31  &  0.000 \\          
  0.008 &      60 &    10.87  &    4.22  &  0.000 \\         
  0.008 &     100 &    14.23  &    3.44  &  0.000 \\
  0.02  &      12 &    11.50  &   18.93  &  0.667 \\    
  0.02  &      15 &    14.21  &   13.34  &  0.655 \\       
  0.02  &      20 &    18.06  &    9.17  &  0.630 \\      
  0.02  &      30 &    12.63  &    6.15  &  0.000 \\       
  0.02  &      40 &     5.35  &    5.14  &  0.000 \\       
  0.02  &      60 &     6.01  &    4.13  &  0.000 \\     
  0.02  &     100 &     7.16  &    3.40  &  0.000 \\ 
  0.05  &      12 &    10.85  &   13.95  &  0.577 \\          
  0.05  &      15 &    13.12  &    9.97  &  0.565 \\           
  0.05  &      20 &    16.05  &    6.99  &  0.539 \\          
  0.05  &      30 &     9.84  &    5.15  &  0.000 \\ 
  0.05  &      40 &     3.63  &    4.36  &  0.000 \\ 
  0.05  &      60 &     4.11  &    3.63  &  0.000 \\       
  0.05  &     100 &     4.22  &    3.11  &  0.000 \\          
\hline
\label{input}
\end{tabular}
\end{table}

In Fig.~\ref{mdot_Z} we plot the evolution of the mass loss rate for
all stellar masses. The mass loss rate depends strongly on the initial
stellar mass and composition. It is clear from the figure that the
lowest mass stars maintain for a long time a low mass loss rate while
the most massive ones evolve rapidly ejecting a large part of their
mass in discrete events.  An important consequence is that a star of
solar metallicity with an initial mass of 100\,M$_{\odot}$ ends its life
with around 7\,M$_{\odot}$, while the evolution of star of
12\,M$_{\odot}$ may be followed for almost 20\,Myr at a very low mass loss
rate that implies that its total mass remains roughly constant.

The mass loss rate is also dependent on the metallicity through the
semi-empirical relation included in the stellar models \citep[see][
for details]{bre93}.  We see these differences in 
Fig.~\ref{mdot_Z} where the evolution for the 6 given metallicities
stars are shown.  The lower the metallicity, the smaller the mass loss
rate and smoother the behavior shown on the mass loss rate evolution.

In Table~\ref{input} we summarize some characteristics of the  stellar
input models: For each metallicity Z, column 1, and initial stellar mass
m$_{*}$, column 2, the final mass m$_{end}$ is listed in column 3, and
the final evolutionary time $t_{end}$ in Myr is in column 4. Column 5
shows the final surface abundance  (in mass fraction) of H, $X_{f,H}$
(which we will use to determine the mass of the Helium core in section 3). 
\begin{figure*}
\resizebox{\hsize}{!}{\includegraphics[angle=0]{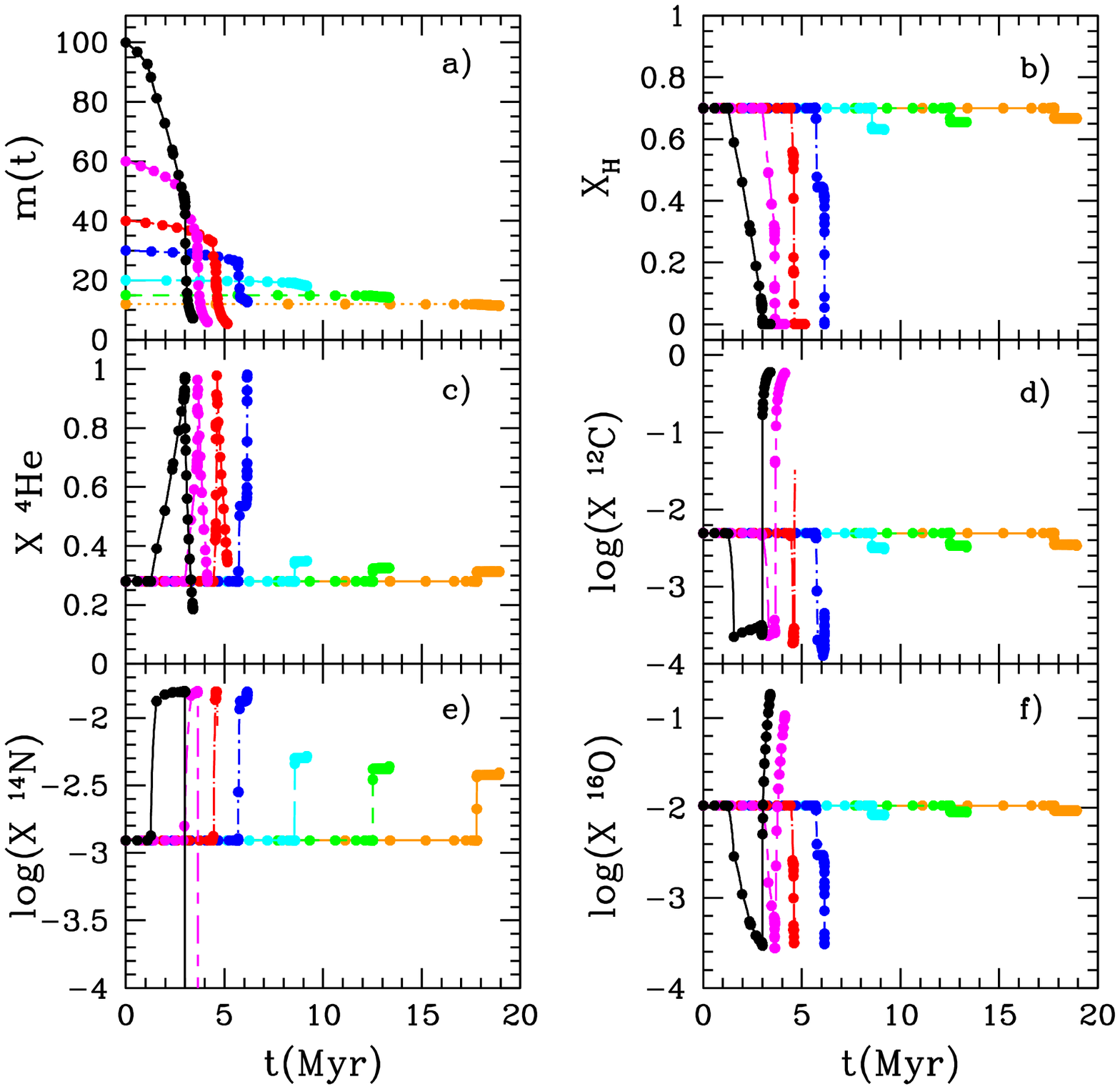}}
\caption{ Input stellar parameters for Z=0.02.  The evolution of the
stellar mass is shown in panel (a) and the evolution of the surface
abundances is shown in panels (b) to (f) for the same initial stellar
masses as labelled in Fig.~1. The dots represent the actual results
from the stellar evolutionary models while the lines are the interpolation in time used in this work.}
\label{stellar_abun}
\end{figure*}

\begin{table*}
\caption{Surface abundances of the stellar models for each mass for Z=0.02. 
Similar complete tables for the six metallicities and seven stellar masses 
will be provided in electronic format}
\begin{tabular}{rrrcccccccc}
Z& N &  m$_{*}$ & t &     $\dt{\rm m}$   &   XH       &  XHe  & XC  & XN   & XO   & m(t) \\
&  & (M$_{\odot}$)& (Myr) & M$_{\odot}.\rm yr^{-1}$&    &            &            &            &            & (M$_{\odot}$)     \\
\hline
0.02 &  7 & 100. &  1.300 &0.364E-04  &0.687E+00 & 0.293E+00  &0.434E-02 & 0.271E-02 & 0.963E-02 &  87.49 \\    
0.02 &  7 & 100. &  1.310 &0.358E-04  &0.684E+00 & 0.296E+00  &0.418E-02 & 0.312E-02 & 0.937E-02 &  87.25 \\    
0.02 &  7 & 100. &  1.320 &0.351E-04  &0.680E+00 & 0.300E+00  &0.402E-02 & 0.353E-02 & 0.911E-02 &  87.00 \\    
0.02 &  7 & 100. &  1.330 &0.345E-04  &0.676E+00 & 0.304E+00  &0.386E-02 & 0.394E-02 & 0.885E-02 &  86.76 \\    
0.02 &  7 & 100. &  1.340 &0.338E-04  &0.672E+00 & 0.308E+00  &0.370E-02 & 0.435E-02 & 0.859E-02 &  86.51 \\    
0.02 &  7 & 100. &  1.350 &0.332E-04  &0.668E+00 & 0.312E+00  &0.354E-02 & 0.476E-02 & 0.832E-02 &  86.27 \\    
0.02 &  7 & 100. &  1.360 &0.325E-04  &0.664E+00 & 0.316E+00  &0.338E-02 & 0.518E-02 & 0.806E-02 &  86.02 \\
\hline
\label{input_stellar} 
\end{tabular}
\end{table*}    

Since each stellar mass table has a different time range and step, we
have performed a linear interpolation to obtain values at the same times for
all masses. The normalized table is provided in electronic format for
all calculated metallicities and masses.  Table~\ref{input_stellar}
shows as an example, a few time steps of the most massive star of
Z=0.02.  It gives for the metallicity of column 1, and for the seven
stellar masses, defined by column 2, the initial mass $m_{*}$ in solar
mass units in column 3, the time in Myr units in column 4, the mass
loss rate, $\dt{m}$, in units of solar mass per year, in column 5, the
elemental abundances of H, $^{4}$He, $^{12}$C, $^{14}$N, and $^{16}$O
\footnote{For the sake of simplicity we write He, C, N and O for
$^{4}$He, $^{12}$C, $^{14}$N, and $^{16}$O along the text.}
as fractions in mass, in columns 6 to 10, and the
stellar mass m(t) at a given time t, in column 11.
 
The evolution of the stellar masses and surface abundances is shown
with dots in Fig.~\ref{stellar_abun}. Each type and color of line
indicate a stellar mass as labelled.  The lines shows the results of
the numerical interpolation used in the following sections.  In panel
a) we see how drastically the stellar mass decreases when $\rm m >
30\,M_{\odot}$ in times as short as 5\,Myr. The evolution of the
stellar surface abundances for H, He in total mass fraction and C, N,
and O, as abundances in mass, $X$, is shown in panels b) to f) of the
same Fig.~\ref{stellar_abun}.  The surface abundances of C, N and O
show a large increase following the start of the stellar winds
revealing the the product of first hydrogen and then helium
burning. This fact combined with a depletion of H in the ejecta means
that, if these abundances were represented as abundances in number,
$12 + \log{(X/H)}$, they would be very high.

To compute the evolution of a stellar cluster ejecta we have assumed
that the cluster stellar mix consists of a coeval population or single
stellar population (SSP) where all stars were created simultaneously and
with the same metallicity.  By using the normalized tables of the
previous paragraph, it is easy to calculate in each time step the
contribution of each star, $m$, weighted by the number of stars in its
mass range, given by the initial mass function $\Phi(m)$. Thus, for
each element i and each time $t$:

\begin{figure*}
\resizebox{\hsize}{!}{\includegraphics[angle=-90]{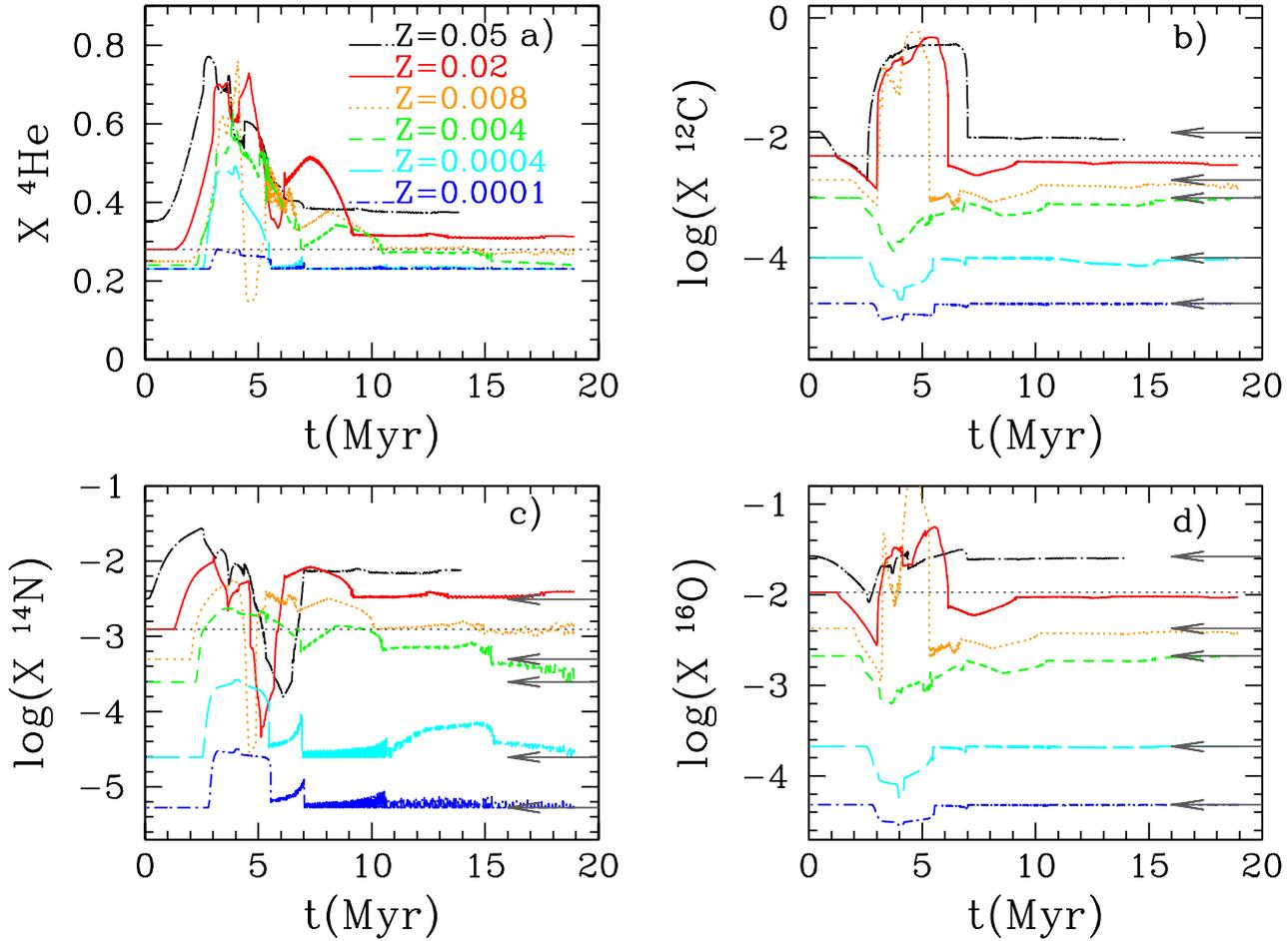}}
\caption{Evolution of the instantaneous abundances in mass for a stellar
cluster due to stellar wind integrated for a  Salpeter IMF: 
a) He, b) C, c) N and d) O for
Z=0,05, 0.02, 0.008, 0.004,0.0004 and 0.0001 with different coded
lines as labelled in panel a. In each panel the dotted grey line marks
the initial abundance for Z=0.02 and the arrows the initial values for
the other metallicities}
\label{abun_Z}
\end{figure*}

\begin{equation}
m_{ej,i}(t)=\int_{mlow}^{mup}\int_{\Delta t}e_{z,i}(m,t') \phi(m) dm dt'
\end{equation}
where
\begin{equation}
e_{z,i}(m,t')=XS_{i}(m,t')\dot{m}(t')
\end{equation}

$XS_{i}(m,t)$ is the surface abundance of each element $i$ and
$\dt{m}(t)$  is the mass loss rate for each stellar mass $m$ in every time
$t$.

We have performed the calculations for 6 different 
initial mass functions: 1) A \citet{sal55} law,
$\Phi(m)\propto m^{-x}$, with an exponent $x= -2.35$ (hereinafter SAL), and 5
others from: 2) \citet[hereinafter MIL][]{mil79}, 3) 
\citet[hereinafter FER]{fer90}, 4) \citet[hereinafter KRO][]{kro01}, 
and 5) \citet[hereinafter CHA][]{cha03}; 
all of them with limits m$_{low}= 0.15$ and m$_{max}=$ 
100\,\Msun\  and 6) a Salpeter law with limits m$_{low}= 1.00$ and m$_{max}=$
100\,\Msun\  as the one used in {\sc{STB99}}.  
Our tables are therefore calculated for all these IMFs, but in the next figures
only Salpeter results are shown since it is a widely used IMF.

The integration is done for the whole mass range of the IMF in each
time step.To integrate in time we have chosen a time step $\delta t=$
0.01\,Myr (small enough to follow the rapid evolution of the mass loss
process without losing any phase). To integrate in mass it is
necessary to be available a grid with a wide range of masses. So, we
have performed a careful interpolation in mass using the 7 existing
tables.  The method to obtain the mass loss rate for any mass value it
is not straightforward, since it shows abrupt changes in small time
scales.  So we have used the tables of the mass loss rate 
to calculate the actual mass m(t) in any time with high accuracy. 
Then we have interpolated between these values for obtaining a new curve for
each mass $m$, and finally we compute the mass loss rate from this
$m(t)$. To calculate the elemental abundances, we have taken
into account the different phases of each element abundance, interpolating
between two known masses to obtain the points limiting these phases for 
each mass $m$. All computations were done for a SSP cluster with a total mass
in stars of 10$^{6}$\,\Msun \footnote{It is necessary to take into
account that we will give averaged values for this stellar cluster
mass. For stellar clusters less massive than 10$\rm ^{4}\,M_{\odot}$,
the stochastic effects over the initial mass function are important
\citep{cer04,cer06} and they may change the stellar mass distribution
and the corresponding results compared with those we obtain.}.

\begin{figure*}
\resizebox{\hsize}{!}{\includegraphics[angle=-90]{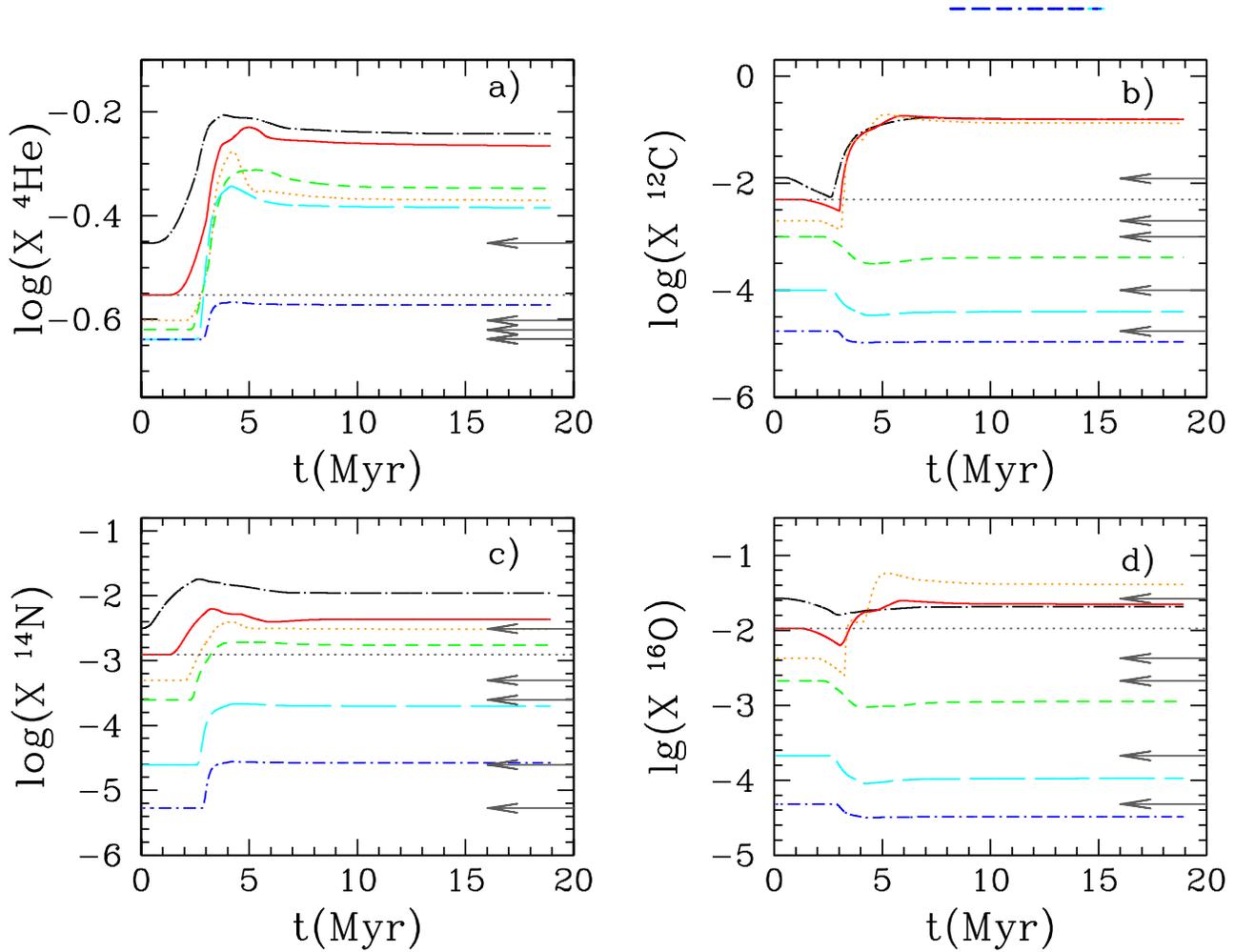}}
\caption{Evolution of abundances in mass of the accumulated 
masses ejected by a stellar cluster for a) He, b) C,c) N and d) O. 
The  metallicities are shown with the same color codes as in the
previous figure. The dotted line and the arrows have the same meaning than in 
Fig. 3}
\label{acum_Z}
\end{figure*}
\begin{table*}
\caption{Accumulated Masses ejected by a stellar cluster of 10$^{6}$\,\Msun\
during its wind phase for 20\,Myr for Z=0.02 and a  Salpeter IMF. 
The complete tables for all times, metallicities  and IMFs
will be provided in  electronic format}
\begin{tabular}{ccrcccccc} 
 IMF & Z & time & m$_{ej}$ & H & He & C & N & O \\ 
     &   & (\,Myr)& (\Msun)&  (\Msun)& (\Msun) & (\Msun)& (\Msun)& (\Msun)\\ 
\hline \\
SAL&     0.0200 &    0.50 &0.753E+03  &   0.527E+03  &   0.210E+03   &  0.372E+01  &   0.934E+00   &  0.798E+01 \\
SAL&     0.0200 &    0.51 &0.770E+03  &   0.539E+03  &   0.215E+03   &  0.380E+01  &   0.954E+00   &  0.816E+01 \\
SAL&     0.0200 &    0.52 &0.787E+03  &   0.550E+03  &   0.220E+03   &  0.388E+01  &   0.975E+00   &  0.834E+01 \\
SAL&     0.0200 &    0.53 &0.804E+03  &   0.562E+03  &   0.225E+03   &  0.397E+01  &   0.996E+00   &  0.852E+01 \\
SAL&     0.0200 &    0.54 &0.821E+03  &   0.574E+03  &   0.229E+03   &  0.405E+01  &   0.101E+01   &  0.870E+01 \\
SAL&     0.0200 &    0.55 &0.838E+03  &   0.586E+03  &   0.234E+03   &  0.414E+01  &   0.103E+01   &  0.888E+01 \\ 
SAL&     0.0200 &    0.56 &0.855E+03  &   0.598E+03  &   0.239E+03   &  0.422E+01  &   0.106E+01   &  0.906E+01 \\
SAL&     0.0200 &    0.57 &0.872E+03  &   0.610E+03  &   0.244E+03   &  0.430E+01  &   0.108E+01   &  0.924E+01 \\
\hline
\label{wind_ejections_pad}
\end{tabular}
\end{table*}

In Fig.~\ref{abun_Z} we show the resulting evolution of the
instantaneous ejecta abundances $X$ for He, C, N and O in mass
as before.  Initial abundances are the ones included in the
original files for Z=0.0001, 0.0004, 0.004, 0.008, 0.02 and 0.05.
If we take the solar abundances from \citet{asp09}
as reference, which implies a total $Z_{\odot}=0.0142$, the
Z=0.02 abundances, indicated in the figure as dotted grey lines, are
in fact $\sim$ 0.15\,dex higher than the adopted Solar values. If we
analyze the results for Z=0.02 (red lines) we see that cluster ejecta
abundances are not always equal to the initial value.  It is evident that ejecta
abundances show strong variations in the WR stars phase, that may
reach up to two orders of magnitude over the initial value for each
metallicity.  Thus, He and N increase in a first phase, when C and O decrease.
Then N and He decrease again just when C and O increase. 

This effect is strongly dependent on the original metallicity of the
stellar cluster as we may see by comparing with the other metallicity
lines.  The initial values for the non-solar cases are indicated with
arrows around the values, +0.3, -0.4,-0.7, -1.7 and -2.3\,dex.  For the
two models with the lowest metallicity, the ejecta abundances, in
particular C and O, do not differ very much from the initial
values except for the decrease around 4-5 Myr.
 However, for higher initial metallicities, all lines show a
strong sudden increase at around 4.5-5\,Myr reaching values 1 or 2
orders of magnitude larger than the initial value and lasting few
million years.  
\begin{figure}
\resizebox{\hsize}{!}{\includegraphics[angle=-90]{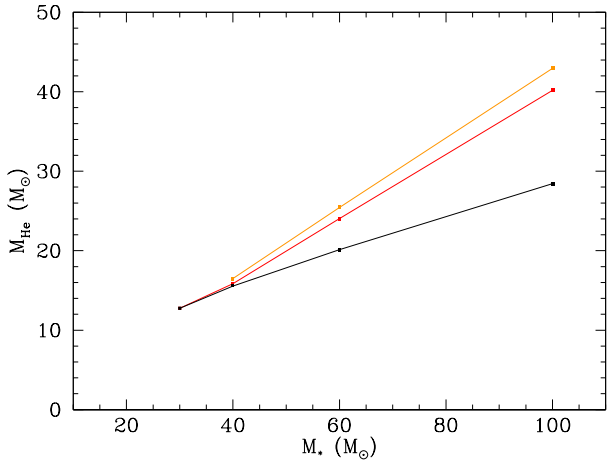}}
\caption{The relation between  the He core mass m$_{He}$, defined
as the stellar mass when the H abundance is zero, and the initial
stellar mass at the main sequence m$_{*}$ for models with m$\ge$ 30\,\Msun.
Each line corresponds to a metallicity as labelled.}
\label{M-MH0}
\end{figure}

\begin{figure} 
\resizebox{\hsize}{!}{\includegraphics[angle=-90]{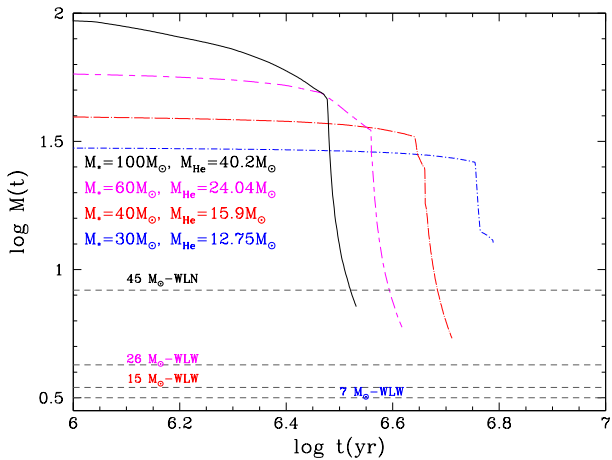}}
\caption{Evolution of the instantaneous mass in the higher mass stellar models.
 The horizontal lines give the WLW93/WLW95 final masses.}
\label{mend}
\end{figure}

For N the behavior is however different. For all models the N
abundance in the ejecta increases at 3\,Myr, even at the lowest
metallicities.  After about 5 to 8\,Myr of evolution, when the cluster
turnoff mass is below 25\,M$_{\odot}$ the ejecta metallicity
asymptotically approaches the initial value for all initial
compositions, except again for N which maintains a higher value than
the initial one. Probably this may be explained by the fact that the
convective envelope dredges up the modified composition of CNO from
the inner parts of the star up to the surface during the Red
Supergiant (RSG) RSG phase, which enhances the N mass fraction,
sligthly reducing C and O abundances.

The resulting accumulated ejected masses for the cluster in every time
step are given in Table \ref{wind_ejections_pad}. For each IMF, column 1,
and metallicity, given in column 2, we give the time step in \,Myr in
column 3, the total ejected mass in column 4, and the accumulated
ejected mass of the different elements in columns 5 to 9 for H, He, C,
N, and O. We show an example here for solar metallicity and a  Salpeter IMF, 
the complete tables for all times, metallicities and IMFs will be provided 
in electronic format.

The evolution of the accumulated ejecta abundances shows in
Fig.~\ref{acum_Z} a sharp increase for Z $\ge 0.008$ due to the mass
loss of massive stars followed by a plateau after the peak of mass
loss associated with the WR stars decline.  For the models with the
lowest metallicities, C and O abundances do not change much in
relation with the initial values, as explained before, decreasing
slightly, simultaneously to the increase of N.  However, for the other
metallicities the increase in C and O abundances is quite large.  On
the other hand, He and N abundances, even at the lowest metallicity
cases, show the sharp increase at 3\,Myr followed by a plateau that
shows an enrichment with respect to the initial value larger than
0.5\,dex in most cases. This increase in the He and N abundance may
have important consequences for measurements of the chemical
composition of galaxies based on abundances estimated using emission
lines from H{\sc ii} regions particularly for the lowest metallicity
regions {\sl i.e.}  those with 12+log(O/H) $\le 7.3$.

These huge variations in the abundances of the ejecta of a stellar
cluster may also be important for the hydrodynamical evolution of the
ISM.  The high metallicity might lead to extremely short cooling times
in the ejecta with important consequences for the subsequent feedback.

\section{Adding the supernova ejecta to the stellar winds}
\begin{table}
\caption{Relation stellar mass-He core mass for Padova stellar tracks}
\begin{tabular}{rrrr}
\hline
         & \multicolumn{3}{c}{Z} \\
         &        0.008 & 0.02      & 0.05 \\
\hline
m$_{*}$  & \multicolumn{3}{c}{M$_{He}$} \\

30        &        \dots   &  12.05    &  9.84 \\
40        &        17.04 &  14.50    & 15.56  \\
60        &        25.44 &  27.80    & 20.13  \\
100       &        45.14 &  42.20    & 28.44  \\
\hline
\end{tabular}
\label{M-MHe}
\end{table}

In this section we include the elements produced by supernova in our
calculations.  Stars more massive than $m\ge 12$\,\Msun\ end their
evolution as core collapse supernovae. New elements are created
and ejected in these events.  Since we assume an IMF were the most
massive star $m_{max}=100\,M_{\odot}$ has a mean lifetime of
$\tau=3.7\,\rm{Myr}$, the ejections are zero before this time. Only
after 3.7\,Myr supernovae begin to contribute to the cluster ejected mass.

In computing the total ejected mass and its composition, an important
effect to take into account is that due to the stellar wind the mass
of a star at the pre-supernova stage is smaller than its main sequence
mass. Thus, the supernova yields for a given star are not those corresponding
its main sequence mass since the star which explodes is less massive. 
To estimate the supernova yields we took as the
supernova progenitor  mass, the mass of each star at the end of its wind phase.  
In practice the models  behave in two different ways  depending on the
wind mass loss rate:

\begin{enumerate}
\item {\sl Small mass loss rate}. Stars with initial masses around
15\,\Msun\ lose only part of their H, therefore $X_{H} > 0$ at all
times. For example stars with $Z=0.02$ and initial masses 12, 15 and
20\,\Msun\ end the winds phase with 11.46, 14.21 and 18.06\,\Msun,
respectively.  In this case, and even more so for smaller abundances,
it is reasonable to assign them the WW95 models in the same way as
PCB98 did. In those cases we simply use the final mass of each star to
select the most appropriate model among the models given by WW95.
\begin{figure*}
\resizebox{\hsize}{!}{\includegraphics[angle=-90]{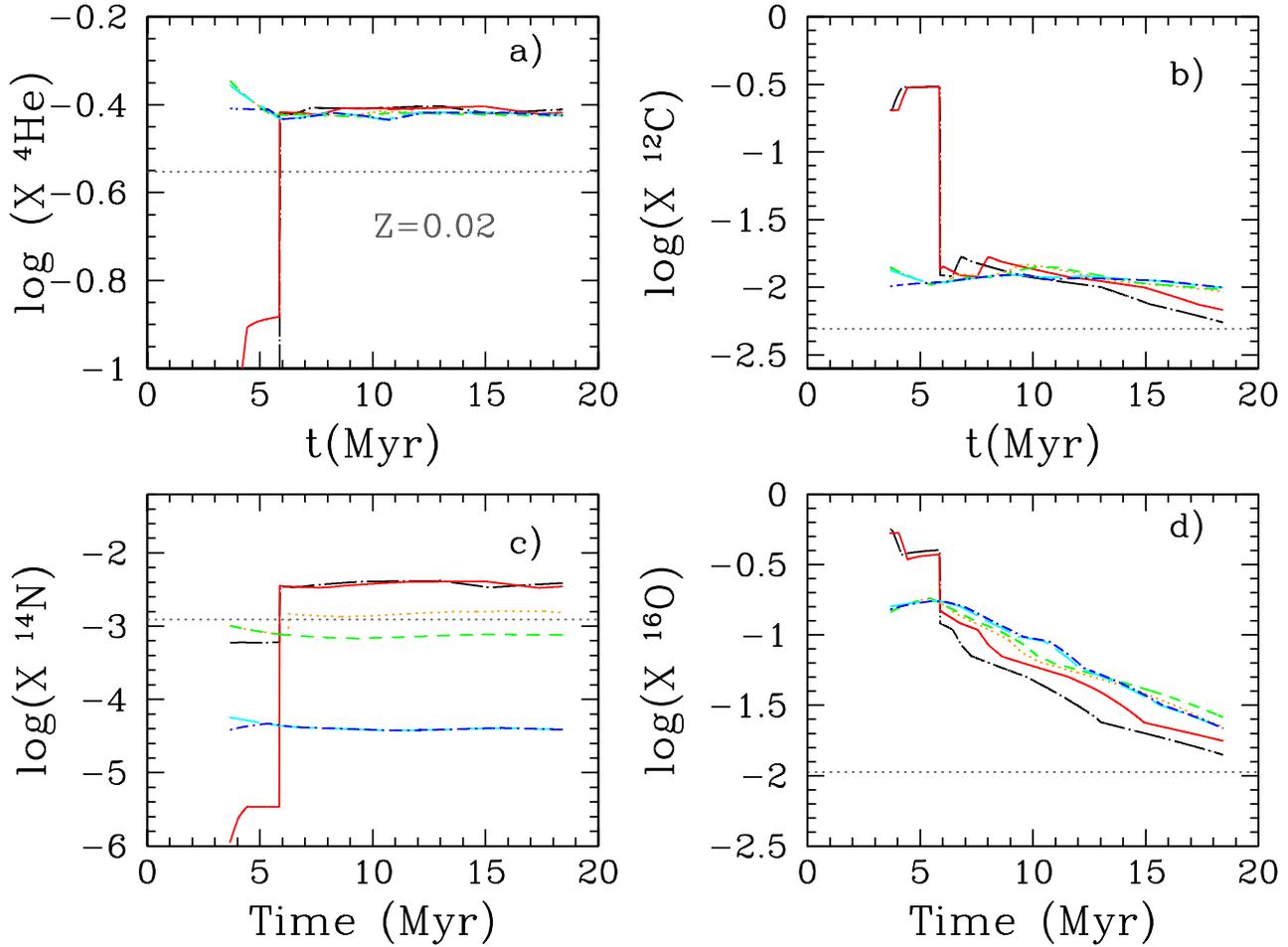}}
\caption{Evolution of the instantaneous values of He, C , N, and O
abundances in mass in the ejecta of a stellar cluster during the
supernova phase. The line codification is as in figure 4.The dotted
grey line marks the abundance Z=0.02 in each panel.}
\label{ejec_supernova}
\end{figure*}
\item {\sl High mass loss rate}. Stars with Z$\ge$ 0.008 and M$\ge
30$\,\Msun  have high mass loss rates and arrive to the end of the
wind phase without H envelope inducing important changes to the supernova
explosion mechanism and ejecta.  For this case we adopted the models
by \citet[][ hereinafter WLW93 and WLW95, respectively]{wlw93,wlw95}.
WLW95 calculated the evolution of stars between 4 and 20\,\Msun\
without H envelope.  Their results are given for a range of masses of
Helium core $\rm m_{He}$, defined as the mass at which the abundance
of H falls to zero.  After H exhaustion, these stars continue losing
mass.  A star of $\rm m_{He}=20$\,\Msun\ loses 16.44\,\Msun, ending
with a mass of 3.55~\,\Msun.  Then, it explodes ejecting other
2.00\,\Msun\ and keeping 1.55\,\Msun\ in the remnant. In similar way a
star of $\rm m_{He}=7.00$\,\Msun\ loses 3.80\,\Msun\ ending with
3.20\,\Msun\ before exploding, ejecting 1.70\,\Msun\ and producing a
remnant of 1.50\,\Msun.

Using the Padova tracks we obtain for the seven original values of
stellar masses, the mass of each star when H is depleted from the
envelope. These values are shown in Table~\ref{M-MHe} for Z$\ge 0.008$
given that for metallicities lower than this the condition $\rm
X_{H}=0$ is never reached.  The relation between the initial mass in
the main sequence m$_{*}$ and the mass of the He core m$_{He}$ is
shown in Fig.~\ref{M-MH0} where we see that the behaviour is quite
smooth with the resulting m$_{He}$ ranging from 10 to 45\,\Msun. The
explosive yields for masses m$_{He}$ between 4 and 20\,\Msun\ are
taken from WLW95. For masses higher than 20\,\Msun, we take the
results from WLW93 where the evolution of WR stars is computed
starting at $\rm X_{H}$=0.50.  For models with initial mass of 60 and
85\,\Msun, (which begin the RSG or Luminous Blue Variable (LBV) phase
with 55.5 and 76.9\,\Msun, respectively), the resulting He core masses
are $\rm m_{He}$= 26.3\,\Msun\ and 45.3\,\Msun, respectively, their
masses at the end of the wind phase being only 4.25 and 8.30\,\Msun.
These models are very similar to our most massive stars of 60 and
100\,\Msun, for which $\rm m(X_{H}$=0.50) are 44.4 and 80\,\Msun, with
$\rm m_{He}=$ 27.8 and 44.9\,\Msun\ and with final masses of 5.93 and
7.16\,\Msun.

In Fig.~\ref{mend} we show the comparison between the evolution of the
most massive stars used here with the final results of WLW93/WLW95.
Note that the final masses in the Padova models for Z=0.02 are slightly larger
than the ones from WLW95/WNL93 with the exception of the most massive
model.  For this case the Padova model reaches 7.17\,\Msun\ at the end
of the evolution while the corresponding model from WLW93 has a
slightly higher value (8.30\,\Msun).
\begin{figure*}
\resizebox{\hsize}{!}{\includegraphics[angle=-90]{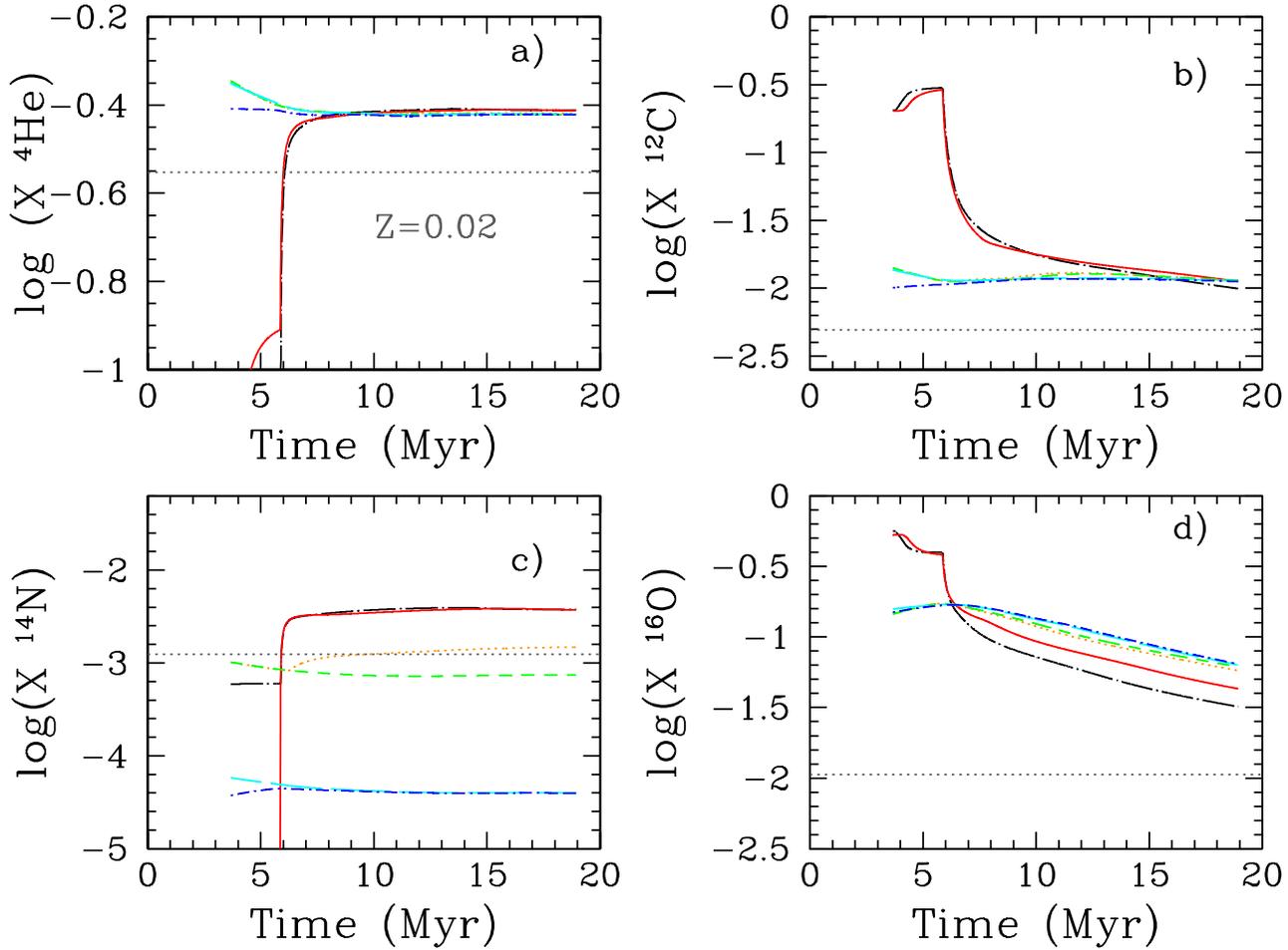}}
\caption{The evolution of the abundances in mass as $\log{X}$ 
of the accumulated ejected mass by a stellar cluster during the supernova phase 
for a) He, b) C, c) N  and d) O  for Z=0.02 and a  Salpeter IMF. 
The dotted  grey line marks the reference abundance  in each panel.}
\label{acum_supernova}
\end{figure*}
\begin{table*}
\caption{Accumulated masses ejected by stars dying as supernova explosions for a
stellar cluster of 10$^{6}\,M_{\odot}$ during the first 20\,Myr 
for Z=0.02 and a  Salpeter IMF. The complete table with 
all times, metallicities and IMFs will be provide in electronic format.}
\begin{tabular}{rrrccccccccc} 
IMF & Z & time & m$_{*}$ & $\rm m_{He}$  & mrem &  mej & H & He & C & N & O \\ 
    &   & (\,Myr)& (M$_{\odot}$)& (M$_{\odot}$) & (M$_{\odot}$)& (M$_{\odot}$)& (M$_{\odot}$) & (M$_{\odot}$)& (M$_{\odot}$)& (M$_{\odot}$) & (M$_{\odot}$)\\ 
\hline 
IMF & 0.02 & 0.37700E+01  &   86.5627  &  34.7581   &  1.5500  & 0.2452E+01  & 0.0000E+00  & 0.1971E+00  & 0.4966E+00  & 0.3056E-05 &  0.1295E+01 \\
IMF & 0.02 & 0.37800E+01  &   85.5161  &  34.3357   &  1.5500  & 0.2698E+01  & 0.0000E+00  & 0.2169E+00  & 0.5465E+00  & 0.3403E-05 &  0.1425E+01 \\   
IMF & 0.02 & 0.37900E+01  &   84.4982  &  33.9250   &  1.5500  & 0.2946E+01  & 0.0000E+00  & 0.2368E+00  & 0.5969E+00  & 0.3762E-05 &  0.1556E+01 \\   
IMF & 0.02 & 0.38000E+01  &   83.5080  &  33.5255   &  1.5500  & 0.3196E+01  & 0.0000E+00  & 0.2570E+00  & 0.6476E+00  & 0.4130E-05 &  0.1689E+01 \\  
IMF & 0.02 & 0.38100E+01  &   82.5443  &  33.1366   &  1.5500  & 0.3448E+01  & 0.0000E+00  & 0.2772E+00  & 0.6987E+00  & 0.4510E-05 &  0.1822E+01 \\ 
IMF & 0.02 & 0.38200E+01  &   81.6059  &  32.7580   &  1.5500  & 0.3702E+01  & 0.0000E+00  & 0.2977E+00  & 0.7502E+00  & 0.4900E-05 &  0.1956E+01 \\ 
IMF & 0.02 & 0.38300E+01  &   80.6921  &  32.3893   &  1.5500  & 0.3957E+01  & 0.0000E+00  & 0.3182E+00  & 0.8021E+00  & 0.5301E-05 &  0.2091E+01 \\   
IMF & 0.02 & 0.38400E+01  &   79.8017  &  32.0300   &  1.5500  & 0.4214E+01  & 0.0000E+00  & 0.3389E+00  & 0.8543E+00  & 0.5713E-05 &  0.2227E+01 \\ 
\hline
\label{supernova_ejections_pad}
\end{tabular}
\end{table*}

Thus, to compute the explosive yields for massive stars with high mass
loss rate, we proceeded as follows: for each time step of our
Table~\ref{wind_ejections_pad}, we calculate the stellar mass that
corresponds to the stellar mean-lifetime $\tau(m)=t$.  Interpolating in
table~\ref{M-MHe} we obtain $\rm m_{He}$, i.e. the mass of the star at
$\rm X_{H}$=0.  We use this value of $\rm m_{He}$ to interpolate with the
adequate value of mass in the explosive yields from WLW95/WLW93 and thus
to calculate the stellar yields that correspond to this star. The yields
are then multiplied by the number of stars given by $\Phi(m)$ for the
initial (zero age main sequence) mass of the star.

Because the production of elements given by WLW93 and WLW95 is
calculated only for stars with Z=0.02 we have assumed that the
relative yields for the other two abundances (Z=0.008 and 0.05) of the
high mass loss rate case are similar. Therefore we have calculated
from the total ejected masses (the new elements and old ones) given by
WLW93 and WLW95, the stellar yields, as new elements ejected by supernova as:
 \begin{equation} 
p_{i}=mej_{i}-(m_{*}-m_{rem})*X_{i,0}
\end{equation}

where $X{i,0}$ are the elemental abundances  for each element $i$
corresponding to Z=0.02,  and $\rm m_{rem}$ is the mass of the remnant.

\begin{figure*}
\subfigure{\includegraphics[angle=-90,width=0.495\textwidth]{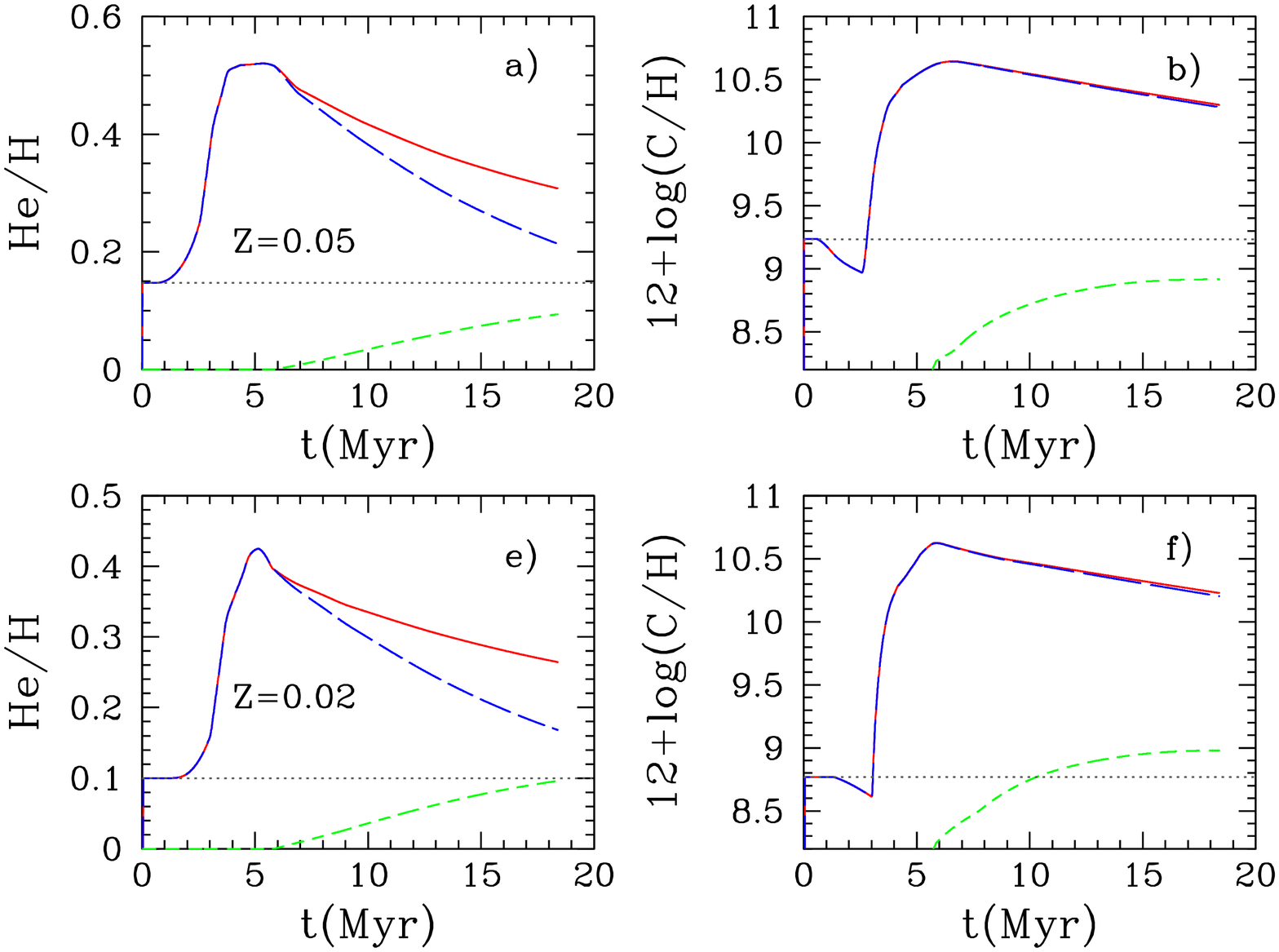}}
\subfigure{\includegraphics[angle=-90,width=0.495\textwidth]{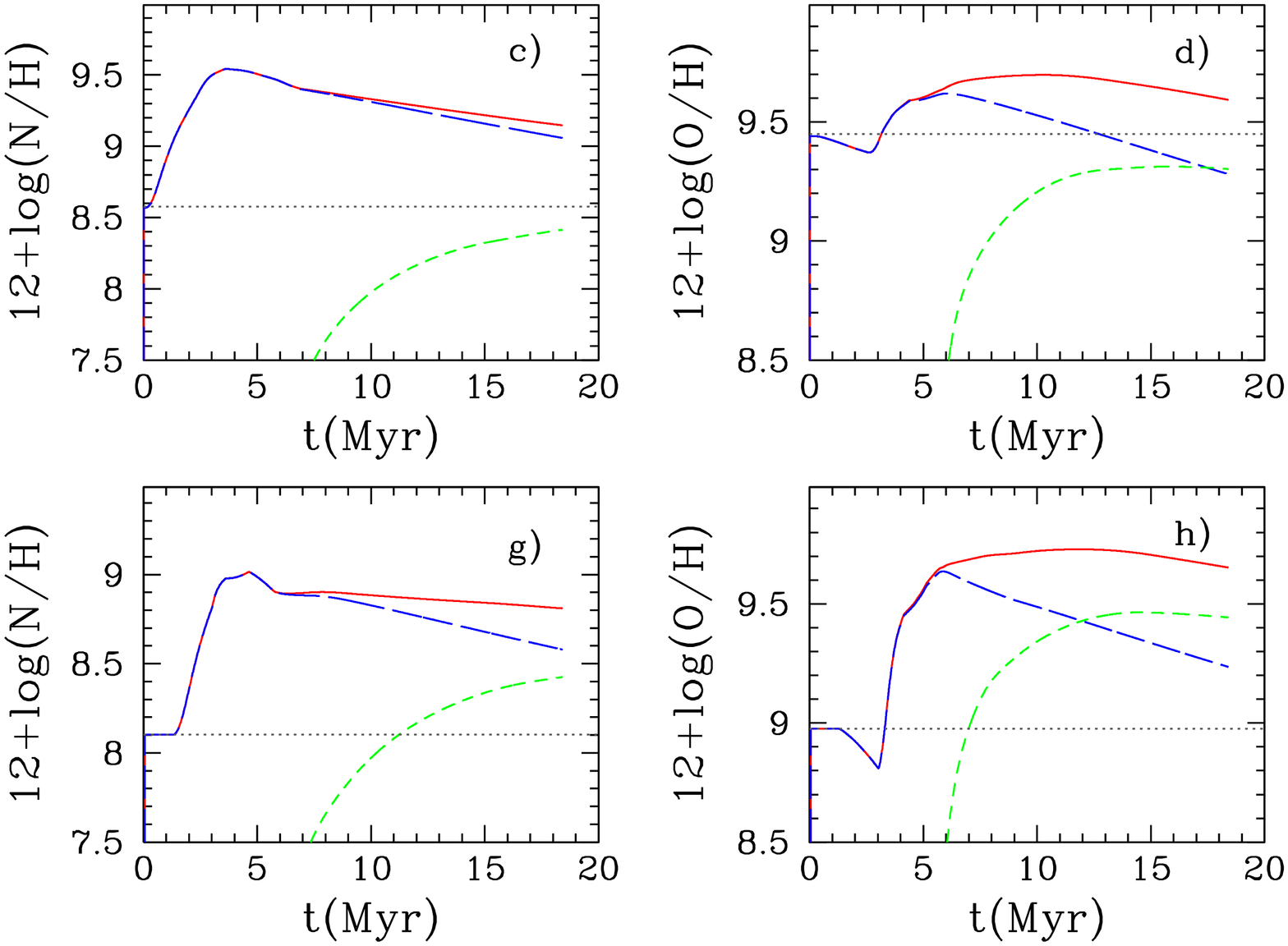}}
\subfigure{\includegraphics[angle=-90,width=0.495\textwidth]{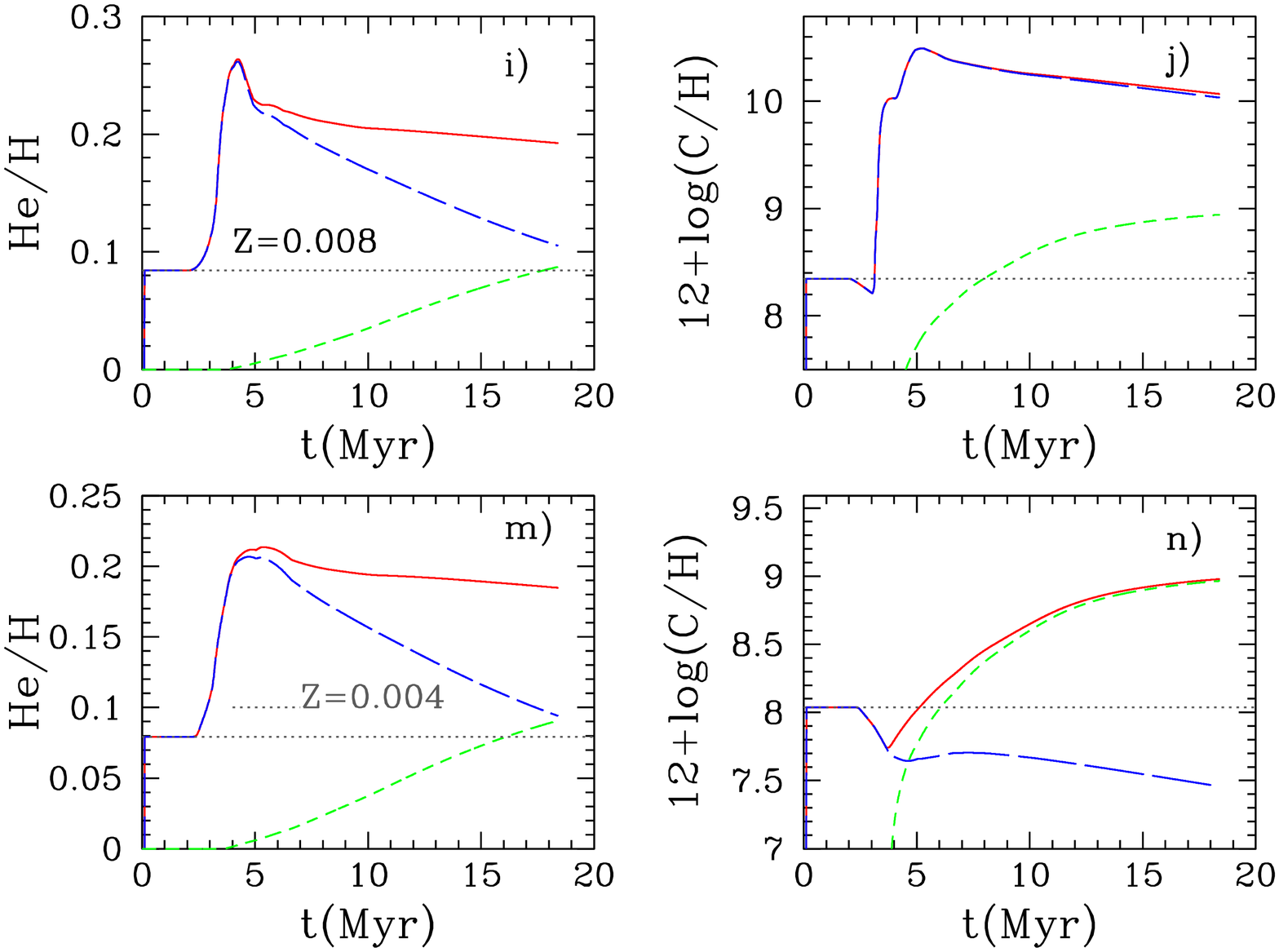}}
\subfigure{\includegraphics[angle=-90,width=0.495\textwidth]{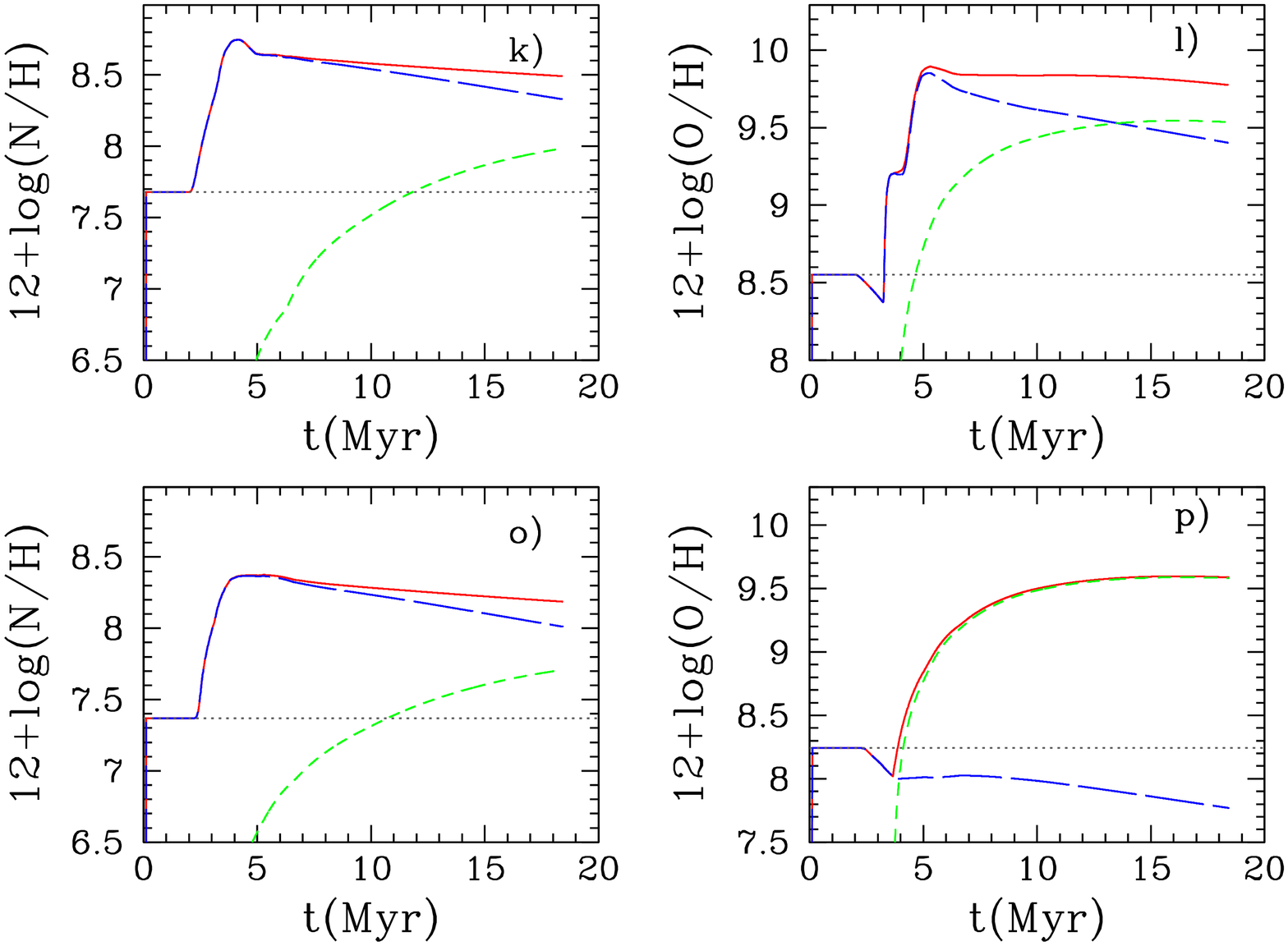}}
\subfigure{\includegraphics[angle=-90,width=0.495\textwidth]{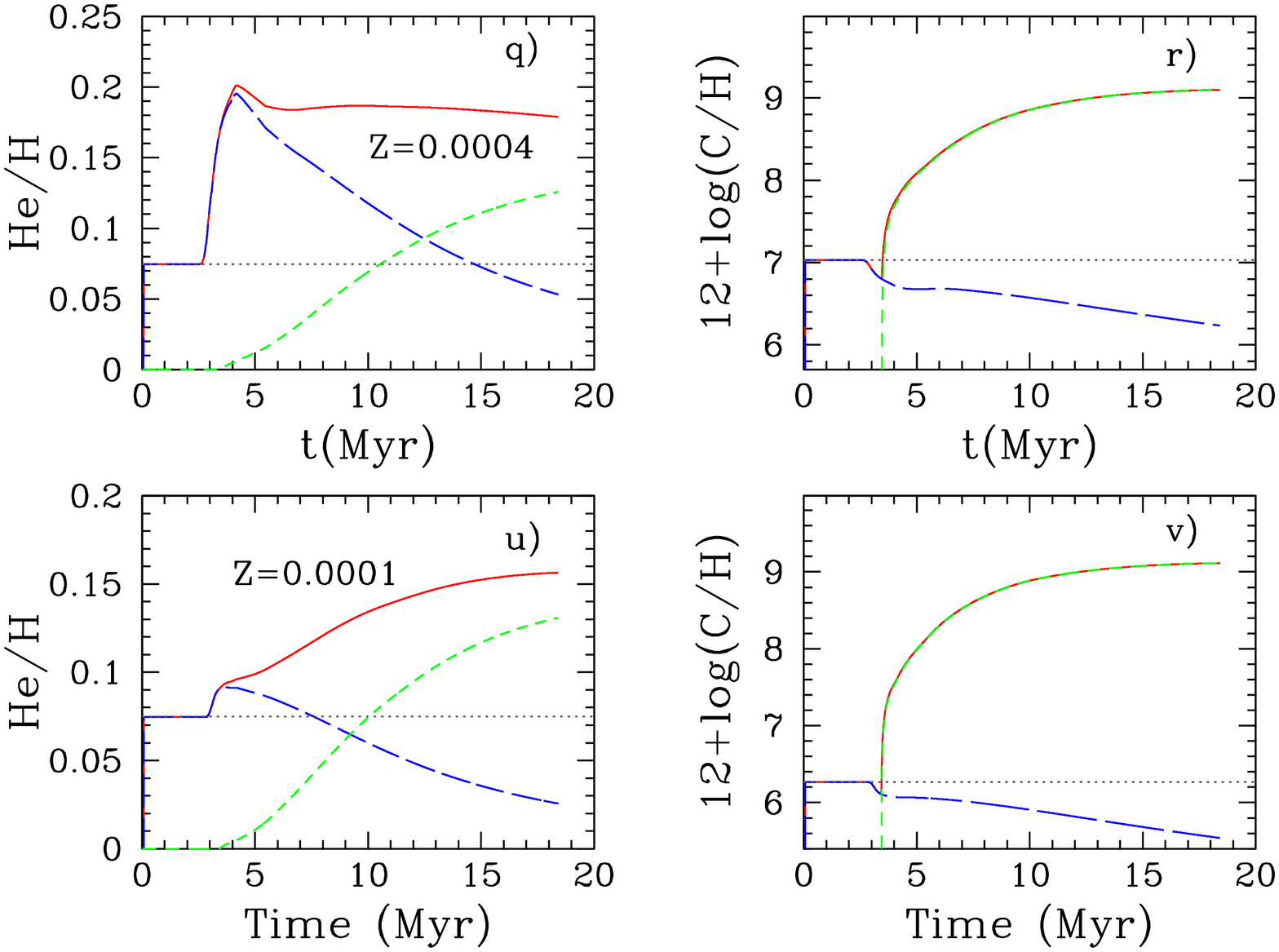}}
\subfigure{\includegraphics[angle=-90,width=0.495\textwidth]{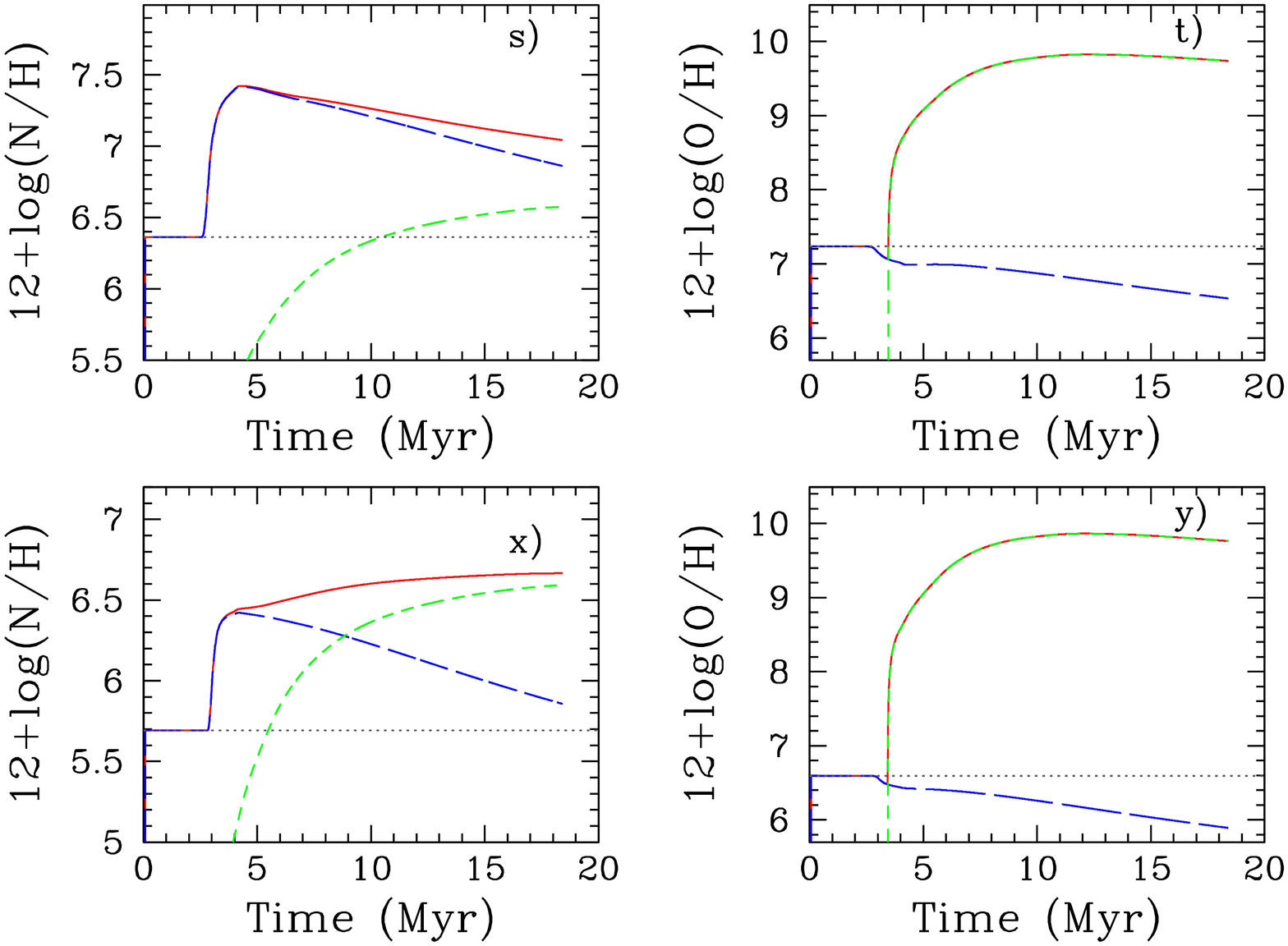}}
\caption{Evolution of the total abundances of the ejecta of a young
stellar cluster with a  Salpeter IMF due to stellar winds and supernova. Stellar wind
contribution: Blue long dash lines. supernova contribution: Short dashed
green lines. Total: Red solid line.The dotted grey line marks the
solar abundance in each panel. The initial abundances are from top to
bottom Z=0,05, 0.02, 0.008, 0.004,0.0004,0.0001. The elemental
abundances are from left to right for He/H, C/H, N/H and O/H. }
\label{tot_abun}
\end{figure*}
Then, we recalculated the ejected masses for each Z as: 
\begin{equation}
mej_{i}=p_{i}+(m_{*}-m_{rem })*X_{i,0}
\end{equation}

for abundances Z=0.008 and 0.05 using the same stellar yield $p_{i}$ 
and  substituting $X_{i,0}$ with the corresponding values.
\end{enumerate}

The instantaneous
mass fraction for He, C, N and O for the supernova ejecta are presented in
Fig.~\ref{ejec_supernova}. Since the ejected mass has no H for supernovaIb/c, we
show all abundances as total mass fraction, as before.  The dotted
line indicates the reference value for Z=0.02.  
Each color indicates a different metallicity with the same coding than
in previous figures 3 and 4.  For the three lowest Z only WW95 has
been used, and so the evolution follows a continuous line. When Z
$\ge$ 0.008 we see the change from the ejecta of supernovaIb (originally WR
stars), following WLW93/WLW95 stellar yields, to the supernovaII from WW95
ejecta as a sharp vertical line. The first ones produce less He and N
and more C and O than the second ones.  This behavior is expected
given that these stars ejected large quantities of He and N before to
explode as supernova, during the stellar winds phase.

Table~\ref{supernova_ejections_pad} gives the accumulated ejected masses by
the supernovaIb/c and supernovaII explosions in each time step. For each
IMF and metallicity given in columns 1 and 2 we show the time in
column 3, the mass of the star which dies in this time in column 4,
the mass of the He core in column 5, the remnant mass in column 6, the
total ejected mass in column 7, and the ejected masses of H to O in
columns 8 to 12.

The evolution of the accumulated or total abundances due to supernova
ejections is illustrated in Fig.~\ref{acum_supernova}. A large increase in
the O and C abundances is clearly seen at 4-6\,Myr corresponding to
the start of the supernova activity.  After that there is a steady decrease
reaching  values around 2 times the initial values after almost 20\,Myr. 
The He and N abundances, however, do not have an important contribution of
supernovaIb and they maintain a constant high level after about 5\,Myr due to
the contribution of supernovaII with lower mass progenitors.

\section{The total abundances of the cluster ejecta}

In this section we show the evolution of the stellar cluster abundances
obtained by adding both contributions, winds and supernova explosions ejections.

In Fig.~\ref{tot_abun} we show the time evolution of the abundances.  
There we plot the contribution of winds by long dashed (blue) lines, the
contribution coming from supernova with short-dashed (green) lines and the
total abundances with solid (red) lines. For He both
contributions are more or less similar at the end of 20\,Myr when
Z$\ge$ 0.004, so the abundance is a factor of two if wind ejections
are considered compared to the usual calculations performed with supernova
productions only. For the two lowest metallicities the ejected masses
are a factor 2 or 3 smaller.

We see that stellar winds produce high abundances of C
and O only for Z$\ge$ 0.004. The level of 12+log(O/H)  doesn't increase 
for Z$\le$ 0.004, while for higher Z it reaches almost two orders
of magnitude larger. N, however, shows higher abundances than
expected for all metallicities, even for the two lowest ones.  
Although these results are not unexpected, since they are due to the
mass loss rate law that depends on Z, these abundances,
not calculated before, may have important consequence over the 
interpretation of observations of H{\sc ii} regions.

When we analyze the same plots for supernovae, we see that now C and O
show very high abundances, compared with the reference values, mainly
for Z$\le $ 0.004, while He and N are roughly in the expected level
for its metallicity. C and O are primary elements, produced directly
from He created in the star. Therefore, their production is expected
to be independent of the initial composition, whereas N is a secondary
element produced in the CNO cycle at the expense of the initial C and
O, so it is reasonable that it scales with the initial abundances.

A summary of these considerations is in Tables~\ref{final_masses_winds}
and ~\ref{final_masses_supernova} where the accumulated masses ejected after
20\,Myr by winds and by supernovae are given. In each one we give for each IMF and each
metallicity the total mass ejected in \Msun\ and then the contributions of each
element, H, He, C, N and O proceeding from stellar winds and from supernovae, 
respectively.

For C the contribution of winds is essential, the supernovae contribution
being a factor of 10 smaller except for the lowest metallicities
(Z=0.0001, 0.0004 and 0.004) for which the ejected masses are insignificant.
Also for N only a small contribution is due to supernovae, at the end of the
evolution, compared with the wind ejecta.  For O the supernovae contribution
is as important as the one from winds, particularly after 10\,Myr of
evolution.  Therefore, the contributions of stellar winds to all
abundances seems to be essential and must be taken into account in the
evolution of stellar clusters and also in the galaxy evolution models
in order to interpret adequately the data.

\begin{figure}
\resizebox{\hsize}{!}{\includegraphics[angle=0]{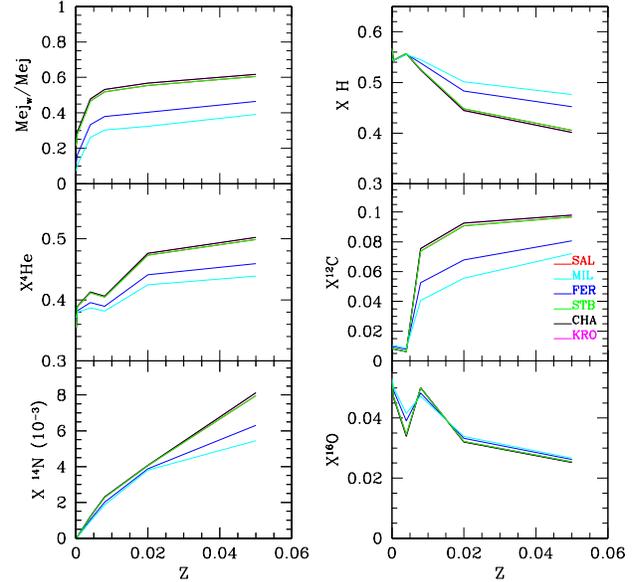}}
\caption{The dependence of the final ejected masses with the metallicity
for different IMFs as labelled}
\label{IMF}
\end{figure}

We show the final results for different IMFs in Fig.~\ref{IMF}.  There
the fraction of mass ejected in the wind phase in panel a) and the
final abundances in panels b) to d) are represented as a function of Z
for the 6 IMFs of this work. Results for MIL and FER show the smallest
abundances of He, C and N, and the highest H values.  The other IMFs
show results very similar, lower than MIL and FER in H, higher for the
other elements. In any case differences among IMFs results are
relatively small.

\begin{table}
\caption{Total masses ejected by stellar winds for a cluster at the
end of the wind phase for all abundances and IMFs.}
\begin{tabular}{rrrrrrrr} 
IMF & Z & mej & H & He & C &  N & O \\ 
    &   & (M$_{\odot}$)&  (M$_{\odot}$)& (M$_{\odot}$)& (M$_{\odot}$) & (M$_{\odot}$)& (M$_{\odot}$) \\ 
\hline 
SAL	&	0.0001	&	12787	&	9363	&	3423	&	0	&	0	&	0	\\
SAL	&	0.0004	&	18080	&	10881	&	7453	&	1	&	4	&	2	\\
SAL	&	0.0040	&	41407	&	24726	&	18598	&	17	&	71	&	47	\\
SAL	&	0.0080	&	52781	&	27930	&	22485	&	6933	&	160	&	2157	\\
SAL	&	0.0200	&	62677	&	24161	&	34006	&	9683	&	269	&	1392	\\
SAL	&	0.0500	&	70809	&	22410	&	40487	&	10853	&	759	&	1446	\\
MIL	&	0.0001	&	2862	&	2106	&	753	&	0	&	0	&	0	\\
MIL	&	0.0004	&	4362	&	2751	&	1646	&	0	&	1	&	1	\\
MIL	&	0.0040	&	13721	&	8515	&	5656	&	7	&	21	&	18	\\
MIL	&	0.0080	&	17733	&	10159	&	6807	&	1920	&	47	&	636	\\
MIL	&	0.0200	&	21829	&	9598	&	10953	&	3253	&	85	&	461	\\
MIL	&	0.0500	&	26913	&	9890	&	13988	&	4567	&	218	&	599	\\
FER	&	0.0001	&	2190	&	1607	&	582	&	0	&	0	&	0	\\
FER	&	0.0004	&	3200	&	1968	&	1268	&	0	&	1	&	0	\\
FER	&	0.0040	&	8553	&	5211	&	3679	&	4	&	14	&	10	\\
FER	&	0.0080	&	10966	&	6058	&	4422	&	1316	&	31	&	422	\\
FER	&	0.0200	&	13278	&	5498	&	6918	&	2021	&	54	&	288	\\
FER	&	0.0500	&	15719	&	5413	&	8538	&	2556	&	146	&	337	\\
MAR	&	0.0001	&	32920	&	20437	&	7472	&	0	&	1	&	1	\\
MAR	&	0.0004	&	46307	&	23750	&	16270	&	2	&	8	&	4	\\
MAR	&	0.0040	&	98511	&	53958	&	40587	&	37	&	155	&	102	\\
MAR	&	0.0080	&	128362	&	60948	&	49078	&	15132	&	349	&	4709	\\
MAR	&	0.0200	&	151042	&	52724	&	74216	&	21128	&	587	&	3038	\\
MAR	&	0.0500	&	169237	&	48903	&	88359	&	23681	&	1656	&	3156	\\
CHA	&	0.0001	&	19514	&	14286	&	5227	&	0	&	1	&	1	\\
CHA	&	0.0004	&	27528	&	16540	&	11381	&	1	&	5	&	3	\\
CHA	&	0.0040	&	62281	&	37132	&	28066	&	25	&	107	&	70	\\
CHA	&	0.0080	&	79380	&	41850	&	33976	&	10484	&	242	&	3254	\\
CHA	&	0.0200	&	94122	&	36074	&	51222	&	14535	&	406	&	2093	\\
CHA	&	0.0500	&	105977	&	33317	&	60835	&	16132	&	1148	&	2155	\\
KRO	&	0.0001	&	18392	&	13464	&	4926	&	0	&	0	&	1	\\
KRO	&	0.0004	&	25946	&	15589	&	10727	&	1	&	5	&	3	\\
KRO	&	0.0040	&	58700	&	34998	&	26452	&	24	&	101	&	66	\\
KRO	&	0.0080	&	74816	&	39444	&	32023	&	9881	&	228	&	3067	\\
KRO	&	0.0200	&	88711	&	34000	&	48277	&	13699	&	383	&	1973	\\
KRO	&	0.0500	&	99885	&	31402	&	57338	&	15205	&	1082	&	2031	\\
STB	&	0.0001	&	25645	&	18777	&	6865	&	0	&	1	&	1	\\
STB	&	0.0004	&	36260	&	21821	&	14947	&	1	&	7	&	4	\\
STB	&	0.0040	&	83041	&	49587	&	37297	&	34	&	142	&	94	\\
STB	&	0.0080	&	105851	&	56013	&	45094	&	13903	&	321	&	4326	\\
STB	&	0.0200	&	125698	&	48455	&	68198	&	19418	&	540	&	2792	\\
STB	&	0.0500	&	142007	&	44943	&	81196	&	21766	&	1521	&	2900	\\
\hline
\label{final_masses_winds}
\end{tabular}
\end{table}

\begin{table}
\caption{Total masses ejected by supernova for a  cluster at the end of the supernova phase for 
each computed metallicity.}
\begin{tabular}{rrrrrrrr}
IMF & Z & Mej & H & He & C & N & O \\ 
    &   & (M$_{\odot}$)&   (M$_{\odot}$)& (M$_{\odot}$)& (M$_{\odot}$) & (M$_{\odot}$)& (M$_{\odot}$)\\ 
\hline 
SAL	&	0.0001	&	46230	&	24090	&	17510	&	521	&	2	&	3089	\\
SAL	&	0.0004	&	46520	&	24230	&	17670	&	529	&	2	&	3058	\\
SAL	&	0.0040	&	47510	&	24800	&	17980	&	549	&	35	&	3018	\\
SAL	&	0.0080	&	49020	&	25610	&	18690	&	563	&	72	&	2934	\\
SAL	&	0.0200	&	50460	&	26510	&	19530	&	581	&	189	&	2246	\\
SAL	&	0.0500	&	46240	&	25050	&	17870	&	468	&	173	&	1522	\\
MIL	&	0.0001	&	36540	&	19470	&	13850	&	410	&	1	&	2093	\\
MIL	&	0.0004	&	36860	&	19650	&	13990	&	416	&	1	&	2076	\\
MIL	&	0.0040	&	38750	&	20630	&	14650	&	443	&	29	&	2164	\\
MIL	&	0.0080	&	40880	&	21760	&	15570	&	463	&	62	&	2138	\\
MIL	&	0.0200	&	45560	&	24180	&	17660	&	492	&	171	&	1821	\\
MIL	&	0.0500	&	41880	&	22880	&	16200	&	390	&	157	&	1236	\\
FER	&	0.0001	&	16310	&	8635	&	6179	&	183	&	1	&	978	\\
FER	&	0.0004	&	16440	&	8706	&	6238	&	186	&	1	&	969	\\
FER	&	0.0040	&	17150	&	9078	&	6484	&	196	&	13	&	994	\\
FER	&	0.0080	&	17990	&	9531	&	6855	&	205	&	27	&	977	\\
FER	&	0.0200	&	19690	&	10420	&	7626	&	216	&	74	&	808	\\
FER	&	0.0500	&	18080	&	9860	&	6991	&	172	&	68	&	548	\\
MAR	&	0.0001	&	100900	&	52600	&	38240	&	1138	&	4	&	6752	\\
MAR	&	0.0004	&	101600	&	52900	&	38590	&	1156	&	4	&	6687	\\
MAR	&	0.0040	&	103700	&	54120	&	39270	&	1198	&	77	&	6596	\\
MAR	&	0.0080	&	107000	&	55890	&	40800	&	1230	&	158	&	6410	\\
MAR	&	0.0200	&	110100	&	57830	&	42610	&	1269	&	412	&	4903	\\
MAR	&	0.0500	&	100900	&	54660	&	38990	&	1022	&	377	&	3323	\\
CHA	&	0.0001	&	66310	&	34510	&	25120	&	748	&	3	&	4476	\\
CHA	&	0.0004	&	66720	&	34700	&	25340	&	759	&	3	&	4430	\\
CHA	&	0.0040	&	68000	&	35440	&	25750	&	786	&	51	&	4357	\\
CHA	&	0.0080	&	70060	&	36550	&	26710	&	806	&	103	&	4231	\\
CHA	&	0.0200	&	71700	&	37640	&	27750	&	830	&	269	&	3215	\\
CHA	&	0.0500	&	65690	&	35570	&	25380	&	670	&	246	&	2179	\\
KRO	&	0.0001	&	62490	&	32520	&	23670	&	705	&	2	&	4218	\\
KRO	&	0.0004	&	62890	&	32700	&	23880	&	716	&	3	&	4176	\\
KRO	&	0.0040	&	64090	&	33400	&	24270	&	741	&	48	&	4107	\\
KRO	&	0.0080	&	66040	&	34450	&	25180	&	760	&	97	&	3988	\\
KRO	&	0.0200	&	67580	&	35470	&	26150	&	782	&	253	&	3031	\\
KRO	&	0.0500	&	61920	&	33520	&	23920	&	631	&	232	&	2054	\\
STB	&	0.0001	&	92710	&	48320	&	35120	&	1046	&	4	&	6195	\\
STB	&	0.0004	&	93300	&	48600	&	35430	&	1061	&	4	&	6133	\\
STB	&	0.0040	&	95280	&	49730	&	36070	&	1100	&	71	&	6052	\\
STB	&	0.0080	&	98310	&	51360	&	37480	&	1130	&	145	&	5883	\\
STB	&	0.0200	&	101200	&	53160	&	39170	&	1165	&	379	&	4504	\\
STB	&	0.0500	&	92740	&	50240	&	35840	&	939	&	347	&	3053	\\
\hline
\label{final_masses_supernova}
\end{tabular}
\end{table}
\section{Conclusions}

We have computed the evolution of the total mass
ejected and the elemental abundances of He, C, N and O for young
stellar clusters of 10$^{6}$ \,\Msun\ for 6 IMFs and for 6 different 
initial metallicities.

Both stellar wind and supernova contribution to the ejecta are
included in the computations. The supernova contribution include the classical
gravitational collapse from \cite{woo95} for low mass progenitors plus
the yields calculated by \cite{wlw93} and \cite{wlw95} for stars more massive
than 30\,\Msun\  and with Z$\ge 0.008$ that reach the supernova stage after
completely losing their H envelope.

The abundances of the ejecta obtained by adding both contributions
show important differences in comparison with the standard method
which uses only supernovae yields without taking into account the
stellar wind yields and/or the difference in the supernovae yields resulting from the huge mass loss affecting the pre-supernova evolution.

In particular:
\begin{itemize}
\item The composition of the ejected matter is determined mostly by supernova
at low metallicities and by stellar winds at around Solar metallicities.

\item The total mass ejected by stellar winds ranges from about 1\% of
the initial cluster mass for the lowest metallicity model to about 6\%
for the $\sim$~Solar abundance ones of the total mass of the cluster
for a  Salpeter IMF.

\item The total mass ejected by supernova is $\sim$5\% of the total mass
of the cluster for all initial metallicities.

\item  At high metallicities the proportion of the mass ejected by the winds phase is around
40-60\% of the total ejecta.

\item There is a large increase in the abundance of He, C, O and N
after 2.5\,Myr with O and  C abundances being the most extreme. The O
abundance jumps almost two orders of magnitude between 2.5 and 4\,Myr in
our lowest metallicity model and about 3 times for the solar abundance
model.  Between 2 and 3\,Myr, the C abundance increases between 10 and
30 times its initial value depending on the initial abundance.

\item He and N show more moderate jumps than C and O in their abundance
between 2.5 and 4\,Myr. He abundance increases almost 3 times for the
solar value models and about 2 times for Z$=0.0004$. On the other hand
N shows jumps of about 5 times for all abundances.  For cluster ages $t <$ 10 Myr, He and N enrichment is mainly due to the stellar winds.

\end{itemize}

These huge variations in the abundances of the ejecta of a stellar
cluster can have a profound effect in the hydrodynamical evolution of the
ISM.  The high metallicity of the young cluster ejecta will lead to extremely short cooling times
 with important consequences for the subsequent feedback.

On the other hand, one should keep in mind the main shortcomings and
uncertainties of our models:

Because present stellar evolutionary models with mass loss do not
include the supernova phase and supernova models do not cover a range in initial
abundances, our adopted values for the supernova yields of the most massive
stars are only approximate.

Binary evolution is not considered. It is known that in young massive
clusters perhaps all massive stars are in binary or multiple systems,
but it is not clear how the presence of a companion would affect the
properties of wind of a massive star.

The effect of stellar rotation is not included in the stellar
evolution models we have used. Again, as in the case of binaries, is
not clear at this stage how rotation would affect the wind of a
massive star.

In spite of these warnings, our models should be useful for the
interpretation of the evolution of the ISM in star forming galaxies.
The resulting tables are available in electronic format.

\section{Acknowledgments}

This work has been partially supported by DGICYT grant
AYA2010--21887--C04--02.  Also, by the Comunidad de Madrid under grant
CAM S2009/ESP-1496 (AstroMadrid) and by the Spanish MICINN under the
Consolider-Ingenio 2010 Program grant CSD2006-00070: First Science
with the GTC \footnote{http://www.iac.es/consolider-ingenio-gtc}.  RT
is grateful to the Mexican Research Council (CONACYT) for supporting
this research under grants CB-2006-49847, CB-2007-01-84746 and
CB-2008-103365-F.  We would like to thank Elena Terlevich and the
referee Rafael Hirschi for for many suggestions that greatly improved
this paper.  Dedicated to the memory of M.F.B.

\end{document}